# Differentially Private Exponential Random Graphs

Vishesh Karwa and Aleksandra B. Slavković and Pavel Krivitsky

**Abstract:** We propose methods to release and analyze synthetic graphs in order to protect privacy of individual relationships captured by the social network. Proposed techniques aim at fitting and estimating a wide class of exponential random graph models (ERGMs) in a differentially private manner, and thus offer rigorous privacy guarantees. More specifically, we use the randomized response mechanism to release networks under $\epsilon$-edge differential privacy. To maintain utility for statistical inference, treating the original graph as missing, we propose a way to use likelihood based inference and Markov chain Monte Carlo (MCMC) techniques to fit ERGMs to the produced synthetic networks. We demonstrate the usefulness of the proposed techniques on a real data example.

## 1. Introduction

Social networks are a prominent source of data for researchers in economics, epidemiology, sociology and many other disciplines and have sparked a flurry of research in statistical methodology for network analysis. In particular, the exponential random graph models (ERGMs) are a very popular modeling framework for analyzing social network data, e.g., see Goodreau, Kitts and Morris (2009), Robins et al. (2007), Goldenberg et al. (2010). While the social benefits of analyzing these data are significant, their release can be devastating to the privacy of individuals and organizations. For example in a famous study by Bearman, Moody and Stovel (2004), researchers analyzed a social network of high school students to study their romantic relationships, and more broadly to understand the structure of human sexual networks. However, such network data are typically only protected via naive anonymization schemes (e.g., by removing the basic identifiers such as name, social security number, etc.), which have been shown to fail and can lead to disclosure of individual relationships or characteristics associated with the released network (for more specific examples, see Narayanan and Shmatikov (2009) and Backstrom, Dwork and Kleinberg (2007)).

In this paper, we develop techniques to provide protection to relationship information while allowing for a valid statistical analysis of the data. We use *edge differential privacy* as a model for measuring privacy risks, and develop inference procedures for analyzing networks using the exponential random graph models.

## 2. Past work on privately estimating ERGMs

Our work is the first to develop techniques for actually fitting and estimating a wide class of ERGMs in a differentially private manner. Previous studies on inferring ERGMs in a private manner have only focused on releasing summary statistics that correspond to sufficient statistics of ERGMs. For example, Karwa et al. (2011) use the smooth sensitivity framework of Dwork et al. (2006a) to add noise and release subgraph counts such as number of k-triangles and k-stars. These subgraph counts are sufficient statistics for a wide class of exponential random graph models, see for example Hunter, Goodreau and Handcock (2008). Hay et al. (2009) propose an algorithm for releasing the degree partition of a graph using the Laplace mechanism. They use post-processing techniques to reduces the $L_2$ error between the true and the released degree distribution.

Most of the previous studies dealing with private release of network data fall short of demonstrating how to perform valid statistical inference using the noisy statistics, which is a non-trivial task. They typically advocate using the noisy statistics as is for inference, sometimes followed by some form of post-processing, ignoring the noise addition process. It has been well established in statistical literature that ignoring the noise addition process can lead to inconsistent and biased estimates, see for example, Carroll et al. (2012) and Fuller (2009). Moreover, even if we are to proceed naively by ignoring the noise addition process and pretend that the noisy statistics are the true sufficient statistics, we often cannot perform inference using existing estimation procedures. This is because many estimation procedures may fail to converge or may give meaningless results.

Fienberg and Slavković (2010), for example, show that maximum likelihood estimators (MLEs) for log-linear models of contingency tables do not exist when sufficient statistics are released using a generalization of mechanism proposed by Barak et al. (2007). Karwa and Slavkovic (2012a,b), demonstrate that the MLE may not exist when one uses Laplace mechanism and naive post-processing techniques for releasing degree sequences of random graphs, and present new algorithms to release graphical degree sequences which ensure that the MLE of the $\beta$ model exists; degree sequences are sufficient statistics of a class of ERGMs known as $\beta$ model. Furthermore, building on the work of Karwa et al. (2011), Karwa and Slavkovic (2012a,b) construct an asymptotically consistent and differentially private estimator of the $\beta$ model. The main technique relies on projecting the noisy sufficient statistics onto the lattice points of the marginal polytope corresponding to the $\beta$ model. Marginal polytopes are polytopes of sufficient statistics and existence of MLE is directly tied to the structure of these polytopes. However, approach of Karwa and Slavkovic (2012a,b) does not scale to more general ERGMs as the corresponding marginal polytopes are not well understood (Engström and Norén, 2010).

In this paper, we take a principled approach, rooted in likelihood theory, to perform inference from data released by privacy preserving mechanisms. Our key idea is to release network data using a differentially private mechanism and estimate the parameters of ERGMs by taking into account the privacy mechanism. Thus, let $X = x$ be the data that requires protection and let $P(X; \theta)$ be a model one is interested in fitting. Privacy preserving mechanisms can be modeled as $P(Y|X = x, \gamma)$, i.e., the released data $y$ is a sample from $P(Y|X = x, \gamma)$ whose parameters $\gamma$ of the privacy mechanism are publicly known. Most of the current work advocates on using $P(y; \theta)$ for inference, ignoring the privacy mechanism. In some cases, $y$ is post-processed to minimize some form of distance from $x$. As noted earlier, using $y$ directly can lead to invalid inferences. Declaring the original data $x$ as missing, we develop methods that take the privacy mechanism into account. Thus we use the likelihood $P(Y; \theta, \gamma) = \sum_x P(Y|X, \gamma) P(X; \theta)$ for inference. This approach offers both the improved accuracy in estimation of $\theta$ and meaningful estimates of standard errors.

The rest of the paper is organized as follows. In Section 3, we introduce the key definitions of differential privacy and the randomized response mechanism used to release the networks. In Section 4, we develop the inference procedures to analyze networks released by the differentially private mechanism. Section 5 presents the experimental results, and is followed by conclusions in Section 6.

## 3. Differential privacy for graphs and Randomized response

This section introduces the privacy model and the notation used throughout the paper. Let $X$ be an undirected simple graph on $n$ nodes with $m$ edges. A simple undirected graph is a graph with no directed edges, and with no self loops and multiple edges. All the graphs considered in this paper are simple and undirected. Let $\mathcal{X}$ denote the set of all simple graphs on $n$ nodes. The distance between two graphs $X$ and $X'$, is defined as the number of edges on which the graphs differ and is denoted by $\Delta(X, X')$. Each node can have a set of attributes associated with it. We will assume that these attributes are known and public. Thus, we are interested in protecting the relationship information in a graph, which is captured by *edge differential privacy*.

### 3.1. Edge Differential Privacy

Edge differential privacy is defined to protect edges in a graph (or relationships between nodes), as the following definition illustrates.

**Definition 1** (Edge Differential Privacy). *Let $\epsilon > 0$. A randomized algorithm $\mathcal{A}$ is $\epsilon$-edge differentially private if for any two graphs $X$ and $X'$ such that $\Delta(X, X') = 1$ and for any subset $S$ of possible outputs of $\mathcal{A}$,*

$$P(\mathcal{A}(X) \in S) \leq e^\epsilon P(\mathcal{A}(X') \in S).$$

Edge differential privacy (EDP) requires that the distribution of outputs obtained from the algorithm $\mathcal{A}$ on two neighboring graphs (i.e., they differ by one edge) should be close to each other. The parameter $\epsilon$ controls the amount of information leakage. Smaller $\epsilon$ leads to lower information leakage and hence provide stronger privacy protection.

One nice property of differential privacy is that any function of the differentially private algorithm is also differentially private as the following lemma illustrates.

**Lemma 1** (Post-processing Dwork et al. (2006b); Nissim, Raskhodnikova and Smith (2007)). *Let $f$ be an output of a differentially private algorithm applied to a graph $X$ and $g$ be any function whose domain is range of $f$. Then $g(f(X))$ is also differentially private.*

### 3.2. Randomized response for edges

Most differentially private mechanisms perturb the output of a function $f$ applied to a dataset. A basic algorithm for releasing the output of any function $f$ under EDP uses the Laplace Mechanism (e.g., see Dwork et al. (2006a)). This mechanism adds Laplace noise to $f(X)$ proportional to its *global sensitivity*, which is the maximum change in $f$ over neighboring graphs. However, this mechanism is not suitable for releasing synthetic graphs for estimating a large class of ERGMs. This is because in order to use the Laplace Mechanism, we need to fix a set of models apriori and release the corresponding sufficient statistics by estimating their sensitivity.

*Remark:* The techniques presented in this paper can be used for the Laplace Mechanism as well.

An alternative way is to perturb the network directly. We call such algorithms input perturbation algorithms. Randomized response is the simplest example of an input perturbation algorithm where random variables are perturbed by a known probability mechanism. Such designs have been extensively used and studied in the context of surveys when eliciting answers to sensitive questions, e.g., see the monograph by Chaudhuri (1987). It has also been used for statistical disclosure control when releasing data in the form of contingency tables, e.g., see Hout and Heijden (2002). We will use a randomized response mechanism to release dyads of a graph, that is subgraphs of size 2, and generate a synthetic graph.

Let $X$ be a random graph with $n$ nodes, presented by its adjacency matrix. In our setting, the adjacency matrix is a symmetric (0,1)- $n \times n$ matrix with zeros on its diagonal, and it captures if there is an edge or not between the nodes in the graph. We will apply randomized response to each entry of the adjacency matrix of $X$. Algorithm 1 shows how to release a random graph $Y$ from $X$ that is $\epsilon$-edge differentially private. Note that for an undirected graph, we need to release $\frac{n(n-1)}{2}$ binary entries. Let $p_{11}$ be the probability of the same edge appearing in both the graphs $x$ and $y$, and $p_{00}$ if there is no edge in both the graphs.

**Proposition 1** (see for e.g., Ganta, Kasiviswanathan and Smith (2008)). *Algorithm 1 is $\epsilon$-edge differentially private with*

$$\epsilon = log\ max\left\{\frac{p_{00}}{1-p_{11}}, \frac{1-p_{11}}{p_{00}}, \frac{1-p_{00}}{p_{11}}, \frac{p_{11}}{1-p_{00}}\right\}.$$

**Algorithm 1**

1: Let $x = \{x_{ij}\}$ be the vector representation of the adjacency matrix of $X$
2: **for** each dyad $x_{ij}$ **do**
3:    if $x_{ij} = 1$, then $y_{ij} = 1$ with prob $p_{11}$, else $y_{ij} = 0$ with prob $1 - p_{11}$.
4:    If $x_{ij} = 0$, then $y_{ij} = 1$ with prob $1 - p_{00}$, else $y_{ij} = 0$ with prob $p_{00}$.
5:    Let $Y = \{y_{ij}\}$.
6: **end for**
7: **return** $Y$

Proposition 1 shows that Algorithm 1 is differentially private. Note that when any of $p_{00}$ and $p_{11}$ are equal to 0.5, we get $\epsilon = 0$, which provides no information about the original graph and hence offers the strongest possible privacy possible. When either of them are 1, we get $\epsilon = \infty$. When both $p_{00}$ and $p_{11}$ are 1, the algorithm releases the original graph and offers no privacy. When $p_{11} = 1$, the algorithm releases the edges exactly, and when $p_{00} = 1$, the algorithm releases the non-edges exactly. We get a range of $\epsilon$ from 0 to $\infty$ for intermediate values of $p_{00}$ and $p_{11}$. We will assume that the parameters of this algorithm are public, i.e., $p_{00}$ and $p_{11}$ are known, otherwise there are identifiability issues, that is the parameters of the model are not identifiable.

Let $1 - p_{00} = 1 - p_{11} = \pi$, where $\pi$ is the probability of perturbing a dyad. This is a special case of the randomized response mechanism, where we flip the state of each dyad with probability $\pi$. In this case, we get $\epsilon = \log \max \left\{ \frac{\pi}{1-\pi}, \frac{1-\pi}{\pi} \right\}$. Let $X$ be the input graph and $Y$ be the output of the randomized response mechanism. We can think of $Y$ as the output from a noisy sampling mechanism applied to $X$. More precisely, if $p_{00} = p_{11} = \pi$ in Algorithm 1, then the output has the following conditional distribution:

$$P_\pi(Y|X = x) = \prod_{ij} \pi^{\mathbb{I}_{y_{ij} \neq x_{ij}}} (1 - \pi)^{\mathbb{I}_{y_{ij} = x_{ij}}},$$

where $\mathbb{I}_{y_{ij} = x_{ij}}$ takes value 1 if there is the same edge in graphs $x$ and $y$ and zero otherwise. Note that if $\pi = 0.5$, we cannot perform any inference on a model for $X$ as all information in the original data is lost. Moreover, if $\pi > 0.5$, the structure of graph "reverses", i.e., edges become non-edges and vice-versa. Hence to provide non-trivial utility, we set $\pi \in \left(0, \frac{1}{2}\right)$. In this case, Algorithm 1 is $\epsilon$-edge differentially private with $\epsilon = -\log \frac{\pi}{1-\pi}$. Note that for conservative values of $\epsilon$, the algorithm may not provide any utility. For instance, for a target $\epsilon = 1$, $\pi \approxeq 0.27$, meaning with probability 0.27 an edge will be flipped. With $\epsilon = 0.1$, $\pi \approxeq 0.47$, approaching 0.5. As $\pi$ approaches 0.5, the "utility" in the perturbed network approaches 0. The question of what is the correct value of $\epsilon$ remains open, but in our case in order to maintain utility for inference, we do need larger values of $\epsilon$.

## 4. Likelihood based inference of ERGMs from randomized response

Exponential random graph models (ERGMs) for a multivariate distribution of $X$ can be parametrized in the following form according to Besag (1974); Frank

and Strauss (1986):

$$P(X = x; \theta) = \frac{\exp\{\theta \cdot g(x)\}}{c(\theta, \mathcal{X})}, \quad x \in \mathcal{X}. \tag{1}$$

Here $\theta \in \Theta \subset \mathbb{R}^q$ are a vector of parameters, $g(x)$ is a vector of sufficient statistics, and $c(\theta, \mathcal{X})$ is the normalizing constant given by

$$c(\theta, \mathcal{X}) = \sum_{x \in \mathcal{X}} \exp\{\theta \cdot g(x)\}. \tag{2}$$

In absence of any privacy mechanism, $x$ is a fully observed random sample from the model given by equation 1. One of the main challenges in finding the maximum likelihood estimate (MLE) of $\theta$ is that the normalizing constant $c(\theta, \mathcal{X})$ given by equation 2 is intractable due to the sum over all possible graphs in $\mathcal{X}$. A lot of work has been done in estimating the normalizing constant and maximizing the likelihood for estimating ERGMs. For example, Geyer and Thompson (1992) use a stochastic algorithm to compute the MLE for a large class of models that includes ERGMs. They approximate the normalizing constant using a Markov chain Monte Carlo (MCMC) algorithm, and compute the MLE by maximizing the stochastic approximation of the likelihood. More precisely, let $\theta_0 \in \Theta$ be a fixed constant. The ratio of two normalizing constants can be approximated as follows:

$$\frac{c(\theta)}{c(\theta_0)} = \sum_{x' \in \mathcal{X}} \frac{\exp\{\theta \cdot g(x')\}}{c(\theta_0)}$$
$$\approx \frac{1}{M} \sum_{i=1}^{M} \exp\{(\theta_0 - \theta) \cdot g(X_i)\},$$

where $X_1, X_2, \ldots, X_M \stackrel{\text{i.i.d.}}{\sim} P(X = x, \theta_0)$ for some initial guess $\theta_0$. Here $M$ is the number of random graphs sampled. Generally, it is difficult to simulate directly from $P(X = x, \theta_0)$ and we need to resort to MCMC methods to generate the sample. For more details on how to construct Markov chains on the space $\mathcal{X}$ and to sample from ERGMs, see Snijders (2002); Handcock (2003); Morris, Handcock and Hunter (2008).

The above algorithm used to approximate the likelihood can be extended to infer $\theta$ from a private sample $y$. Such extensions were also considered in Handcock et al. (2010) in the context of so called *ignorable* sampling mechanisms for network data, i.e. when $y$ is a sample of the original network $x$. Roughly, ignorable designs are those where the sampling mechanism does not depend on the missing data. Our setting is different because in general, differential privacy mechanisms are not ignorable and depend on the original data. However, as we will see, the MCMC approach of Geyer and Thompson (1992) can be extended to estimate parameters from data released by privacy mechanisms since the parameters of the privacy mechanism are public.

The following discussion is general and applies to a generic privacy mechanism $P_\gamma(Y|X)$ with known $\gamma$. Recall that we wish to estimate $\theta$ using a private sample $y$ obtained from $P_\gamma(Y|X=x)$. A naive approach is to ignore the privacy mechanism and estimate the parameters using the naive likelihood $P(X=y,\theta)$. The correct approach is to include the privacy mechanism in the model and use the full likelihood of $Y$. We can formulate this likelihood by treating the original data $x$ as missing, and summing over all possible values of $x$. Thus, if we let $\hat{\theta}_{mle}(y)$ be the maximum likelihood estimator of $\theta$ obtained from $y$, then

$$\hat{\theta}_{mle}(y) = \underset{\theta \in \Theta}{\mathrm{argmax}}\ L(\theta; y) \tag{3}$$

$$= \underset{\theta \in \Theta}{\mathrm{argmax}} \sum_{x \in \mathcal{X}} P_\gamma(Y=y|X=x) P(X=x;\theta). \tag{4}$$

For our purposes, $P(X=x;\theta)$ is the ERGM we are interested in fitting. $P_\gamma(Y=y|X=x)$ is the privacy mechanism with parameters $\gamma$. In case of the randomized response mechanism of Algorithm 1, $\gamma = \pi$.

With a bit of algebra, we can re-write the likelihood based on $y$ as follows:

$$L(\theta; Y=y, \pi) = \frac{c(\theta|y)}{c(\theta)}$$
$$\text{where } c(\theta|y) = \sum_{x \in \mathcal{X}} e^{\theta \cdot g(x)} P_\gamma(Y=y|X=x).$$

Thus, we need to estimate two intractable constants $c(\theta|y)$ and $c(\theta)$. They can be approximated by using the MCMC technique of Geyer and Thompson (1992) described previously. We need two Markov chains, one for the estimating $c(\theta)$ and the other for estimating $c(\theta|y)$. To estimate the latter constant, the MCMC sample needs to be weighted by the privacy weights $P_\gamma(Y|X)$, which are known. Thus, if $X_1, \ldots X_M \overset{\text{i.i.d.}}{\sim} P(X=x, \theta_0)$

$$\frac{c(\theta|y)}{c(\theta_0)} = \sum_{x' \in \mathcal{X}} \frac{\exp\{\theta \cdot g(x')\} P_\gamma(y|X_i)}{c(\theta_0)}$$
$$\approx \frac{1}{M} \sum_{i=1}^{M} \exp\{(\theta_0 - \theta) \cdot g(X_i)\} P_\gamma(y|X_i),$$

where $M$ is the number of sampled graphs.

Note the key in being able to estimate $c(\theta|y)$ is that the weights $P_\gamma(y|X_i)$ can be computed because the parameters of the privacy mechanism are known. A similar weighting based approach but with the EM-algorithm was proposed by Woo and Slavković (2012) for estimating logistic regression from variables subject to another privacy mechanism known as the Post Randomization Method (PRAM). The standard errors and confidence interval of the parameters can be derived in the usual manner; for details see Morris, Handcock and Hunter (2008).

## 5. Experiments

In this section, we evaluate the proposed differentially private randomized response algorithm to release synthetic networks and estimate the parameters of ERGM using the missing data likelihood. Specifically, we consider a subset of a friendship network collected in the Teenage Friends and Lifestyle Study; see Pearson and Michell (2000) and Michell and Amos (1997). The study records a network of friendships and substance use for a cohort of students in a school in Scotland. In the current study, we used an excerpt of 50 adolescent girls made available online in the *Siena* package (Siena, 2014). The network consists of 50 nodes and 39 edges. There are four covariates associated with each node: *Drug usage* (yes or no), *Smoking status* (yes or no), *Alcohol usage*, (regular or irregular) and *Sport activity* (regular or irregular).

As mentioned earlier, we assume that the covariates associated with each node are available publicly. Our goal is to protect the relationship information in the network $x$. Thus, we release the adjacency matrix of $x$ using the randomized response mechanism of Algorithm 1 for varying values of $\pi$. For each value of $\pi$, we release 10 synthetic networks. For each released network $y$, we fit an ERGM, using two different likelihoods: One that takes the privacy mechanism into account, called the *missing data likelihood*, and the other that ignores the privacy mechanism, called the *naive likelihood*. We fit the following ERGM to the network:

$$P(X;\theta) \propto \exp\{\theta_1 edges + \theta_2 gwesp + \theta_3 popularity + \theta_4 drug + \theta_5 sport + \theta_6 smoke\}. \tag{5}$$

The first three terms in equation 5 capture the network structure of the graph, and the last three terms represent the homophily effect of covariates. The term *edges* measures the number of edges in the network. The term *gwesp* measures the transitive effects in the network, in a weighted manner, and the term *popularity* captures the degree distribution of the network. For more details on these terms, see Morris, Handcock and Hunter (2008). We use the *ergm* package (Hunter et al., 2008) in R (R Core Team, 2014) to fit the models.

We evaluate these methods by measuring the Kullback-Leibler (KL) divergence between the distributions implied by estimates obtained from the private network and the true network. Let $\theta_x$ and $\theta_y$ be two parameter estimates obtained by using the original network $x$ and the private network $y$, respectively. Recall that the KL divergence between the two distributions is given by the following equation:

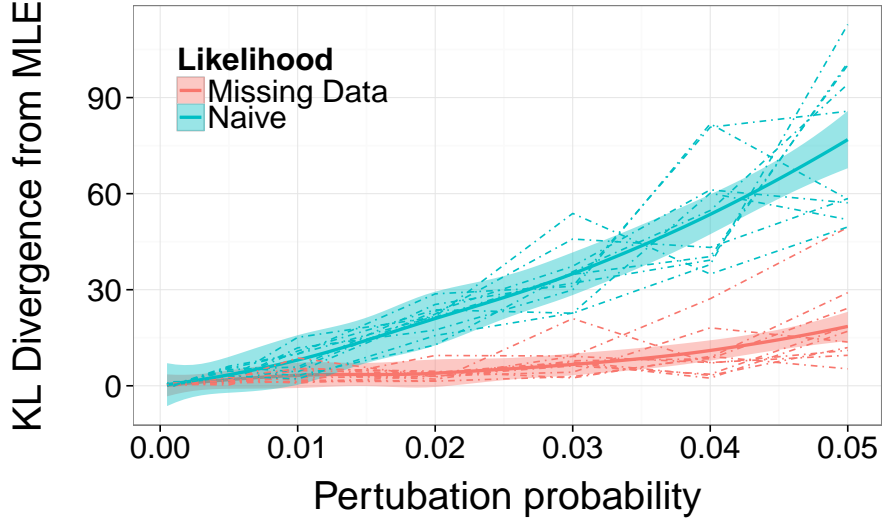

Fig 1: Comparison of the private ERGM models in the Friendship dataset estimated using the missing data likelihood and the naive likelihood. The red line (lower value of KL) represents the KL divergence between the estimates based on the missing data likelihood and the MLE from the original data. The cyan line (higher value of KL) represents the KL divergence between the estimates based on the naive likelihood and the MLE. The $x$-axis represents the perturbation probability $\pi$ used in to release the synthetic network.

$$KL(\theta_x, \theta_y) = E_{\theta_x}\left[\log \frac{P(x, \theta_x)}{P(x, \theta_y)}\right]$$
$$= \sum_{x \in \mathcal{X}} \log\left(\frac{P(x, \theta_x)}{P(x, \theta_y)}\right) P(x, \theta_x)$$
$$= (\theta_x - \theta_y) g(x) + \log \frac{c(\theta_y)}{c(\theta_x)}.$$

The KL divergence can be easily computed using the MCMC techniques described in Section 4; see also Handcock et al. (2010) for more details. Figure 1 shows the plot of the KL divergence between the private and non-private network on the $y$-axis and the perturbation probability $\pi$ on the $x$-axis, for different releases of the synthetic network. The solid line represents the mean KL divergence and the shaded region represents the 99 percent confidence region. The dotted lines show the value of KL divergence for each released dataset. Note that $\epsilon = -\log \frac{\pi}{1-\pi}$, so larger values of $\pi$ imply stronger privacy.

Figure 1 shows that the KL divergence between the private estimate and the non-private estimate increases as $\pi$ increases, thus stronger privacy leads to reduced utility. However, the KL divergence of the naive likelihood increases at a much faster rate when compared to the missing data likelihood. This is true especially for larger values of $\pi$. Thus for strong privacy protection, the missing data likelihood provides estimates that are closer to the non-private estimates when compared to the naive likelihood.

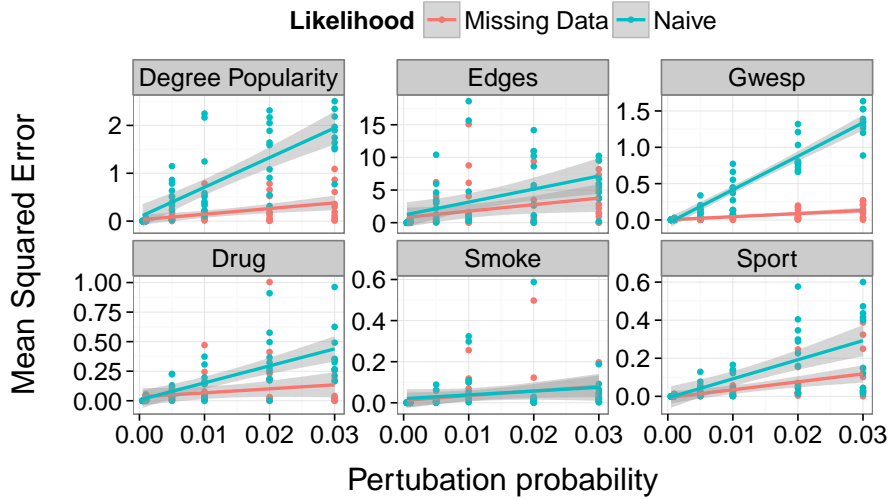

Fig 2: MSE of parameter estimates of the private ERGM models in the Friendship dataset estimated using the missing data likelihood and the naive likelihood. The red line (lower values) represents the MSE of the estimates based on the missing data likelihood. The cyan line (higher value) represents the MSE of the estimates based on the naive likelihood and the MLE. The $x$-axis represents the perturbation probability $\pi$ used in to release the synthetic network.

For a more detailed evaluation of our method, we will look at the relative mean squared error (MSE) of individual parameter estimates. The error is calculated as the mean squared difference between the estimates obtained by the private network and the estimates from the true network. Figure 2 shows a plot of MSE for each parameter in the ERGM model. Note that the first three parameters capture the relational structure of the network and the last three parameters measure the main effects of the nodal covariates *Drug*, *Smoke* and *Sport*. Table 1 shows the estimate of mean percentage relative efficiency of the parameters, i.e., it shows the ratio $\frac{MSE[Missing]}{MSE[Naive]}$ in form of percentage. In the table, values less than 100 favor the proposed missing data estimator.

Figure 2 and Table 1 show that for structural parameters, the MSE of estimates based on missing data likelihood are much smaller when compared to those based on the naive likelihood. This is true specially for larger values of $\pi$.

TABLE 1
Table of Relative efficiency of the estimators, $\frac{MSE[Missing]}{MSE[Naive]}$ for different values $\pi$ of perturbing a dyad.

|  | Parameter | | | | | |
| --- | --- | --- | --- | --- | --- | --- |
| $\pi$ | Popularity | edges | GWESP | Drug | Smoke | Sport |
| 0.005 | 16.5 | 38.1 | 9.8 | 43.4 | 59.2 | 37.7 |
| 0.01 | 26.1 | 80.4 | 15.8 | 74.4 | 69.4 | 31.8 |
| 0.02 | 17.6 | 51.7 | 7.9 | 58.8 | 105.3 | 24.6 |
| 0.03 | 19.9 | 49.3 | 10.2 | 13.2 | 124.3 | 49.4 |

For the parameters related to the homophily effects, the missing data estimates also have lower MSE when compared to the naive estimates. However, the difference is not as drastic as for the structural parameters, especially in case of the node covariate *smoke*. This is due to the fact that the nodal characteristics are assumed to be public, hence the parameter estimates are effected only by the changes in the total number of edges between nodes of the same covariate value. In fact, for the *smoke* parameter, in some cases, as seen in Table 1, the naive estimator seems to do better in terms of MSE. For the structural parameter *gwesp*, the improvement in efficiency is quite substantial when using the missing data likelihood.

## 6. Conclusion

In this paper, we present an $\epsilon$-edge differentially private algorithm to release and estimate exponential random graph models. We release synthetic networks using a randomized response mechanism. By treating the original data as missing, we incorporate the privacy mechanism into the likelihood for estimating the parameters. We show that missing data methodology and MCMC techniques can be directly extended to maximize the likelihood of data released by a differentially private mechanism and more generally any privacy preserving mechanisms.

Simulation studies show that our proposed approach leads to estimates with much lower mean squared errors when compared to those obtained by ignoring the privacy mechanism. Although we advocate the use of missing data and MCMC techniques by analysts who use data obtained from a differentially private mechanism, or more general privacy-preserving mechanisms, they can also be used by data curators to release synthetic graphs for performing preliminary analysis of other models. Indeed, using our techniques, the data curator can fit an ERGM to the data and release synthetic graphs from the ERGM. The utility of the synthetic graphs may depend on the goodness-of-fit of the ERGM chosen by the data curator, and this requires further careful investigation.

In this paper, we assumed that the covariate information is available publicly, which may not always be the case. We are currently working on relaxing this assumption and releasing synthetic graphs that protect both nodal and structural information in a graph. Future investigations will also include evaluating the usefulness of this approach for different sizes and sparsity of networks and other ERGM specifications.

*Acknowledgments.*

This work was supported in part by NSF grant BCS-0941553 to the Department of Statistics, Pennsylvania State University. The authors would like to thank the reviewers for their helpful suggestions.